# Switching teraherz waves with gate-controlled active graphene metamaterials

Seung Hoon Lee[1*], Muhan Choi[2*], Teun-Teun Kim[1*], Seungwoo Lee[1], Ming Liu[3], Xiaobo Yin[3], Hong Kyw Choi[2], Seung S. Lee[1], Choon-Gi Choi[2], Sung-Yool Choi[4], Xiang Zhang[3,5] & Bumki Min[1]

[1]*Department of Mechanical Engineering, Korea Advanced Institute of Science and Technology (KAIST), Daejeon 305-751, Republic of Korea*

[2]*Creative Research Center for Graphene Electronics, Electronics and Telecommunications Research Institute (ETRI), Daejeon 305-700, Republic of Korea*

[3]*Nanoscale Science and Engineering Center, 3112 Etcheverry Hall, University of California, Berkeley, California 94720, USA*

[4]*Department of Electrical Engineering, Korea Advanced Institute of Science and Technology (KAIST), Daejeon 305-751, Republic of Korea*

[5]*Material Science Division, Lawrence Berkeley National Laboratory, 1 Cyclotron Road, Berkeley, California 94720, USA*

[*]These authors contribute equally to this work.

**The extraordinary electronic properties of graphene, such as its continuously gate-variable ambipolar field effect and the resulting steep change in resistivity, provided the main thrusts for the rapid advance of graphene electronics[1]. The gate-controllable electronic properties of graphene provide a route to efficiently**

manipulate the interaction of low-energy photons with massless Dirac fermions, which has recently sparked keen interest in graphene plasmonics[2-10]. However, the electro-optic tuning capability of unpatterned graphene alone is still not strong enough for practical optoelectronic applications due to its nonresonant Drude-like behaviour. Here, we experimentally demonstrate that substantial gate-induced persistent switching and linear modulation of terahertz waves can be achieved in a two-dimensional artificial material, referred to as a metamaterial[11,12], into which an atomically thin, gated two-dimensional graphene layer is integrated. The gate-controllable light-matter interaction in the graphene layer can be greatly enhanced by the strong resonances and the corresponding field enhancement in the metamaterial[13]. Although the thickness of the embedded single-layer graphene is more than 'six' orders of magnitude smaller than the wavelength (< $\lambda/1,000,000$), the one-atom-thick layer, in conjunction with the metamaterial, can modulate both the amplitude of the transmitted wave by up to 90 per cent and its phase by more than 40 degrees at room temperature. More interestingly, the gate-controlled active graphene metamaterials show hysteretic behaviour in the transmission of terahertz waves, especially when fabricated with multilayer graphene, which is indicative of persistent photonic memory effects. The controllable light-matter interaction, as well as the memory effect in the active graphene metamaterials, present immense potential for a myriad of important applications, particularly in active control of terahertz waves in the extreme subwavelength-scale, such as fast terahertz modulators, tunable transformation-optics devices, electrically controllable photonic memory, and reconfigurable terahertz devices.

Terahertz frequencies are situated in the far-infrared spectra of electromagnetic radiation that inherit the richness of photonics and electronics, as well as their weaknesses[14,15]. The development of electrically tunable terahertz semiconductor devices, preferably operating at room temperature, is hindered by the limitation on the change of free carrier density, which leads to inefficient responses to terahertz radiation[16,17]. Within the last decade, considerable effort has been devoted to efficiently modulating terahertz waves, with approaches including utilization of a semiconductor two-dimensional electron gas (2DEG) system[16] and a hybridized metamaterial with a charge carrier injection scheme, such as a Schottky diode[18,19] or a high electron mobility transistor (HEMT)[20]. However, these previous attempts were based on conductivity changes by charge carrier injection on *bulk* substrates, and their electron carrier density is limited to a value of $\sim 1 \times 10^{12}$ cm$^{-2}$. As a possible alternate route, graphene, a truly two-dimensional atomic system, can be employed to circumvent this limit as it allows dramatic modification of the Fermi level and the corresponding charge carrier density by simple electric gating. This charge carrier handling capability, within the limit of dielectric breakdown, is an order of magnitude larger ($\sim 10^{13}$ cm$^{-2}$) than that of conventional 2DEG systems[16,20]. Furthermore, the small effective mass of charge carriers in graphene makes it possible to fully maximize the light-matter interaction in the extreme subwavelength-scale.

Defects in graphene-based devices, such as grain boundaries, adsorbed H$_2$O molecules, and other impurities acting as charge trap sites on dielectric substrates or on a graphene sheet, cause hysteresis in *electronic* transport[21]. This bistable behaviour has opened up a new path to implement graphene-based electronic memory devices using electrochemical modification[22], graphene oxide[23], or ferroelectric materials[24]. On the other hand, a hysteretic change in the *optical* conductivity of graphene, induced by charge carrier retention or delayed response in reaching equilibrium, will provide potential *photonic* memory applications, similar to phase-change memory metamaterials[25]. Here, we present an electrically controllable light-matter interaction in a hybrid material/metamaterial system consisting of artificially constructed meta-atoms and truly two-dimensional carbon atoms. The extraordinary electrical and optical properties of graphene, when enhanced by the strong resonance of meta-atoms, lead to light-matter interaction of an unprecedented degree such that persistent switching and linear modulation of terahertz wave transmission are realized in the extreme subwavelength-scale ($< \lambda/1,000,000$).

The structure of a fully integrated, gate-controlled, active terahertz graphene metamaterial is depicted schematically in Fig. 1a. Functionally, the device is a combination of an array of meta-atoms, an atomically thin graphene layer transferred conformally onto the metamaterial layer, and an array of metallic wire gate electrodes (For details on fabrication of the device, see Methods Summary). In order not to complicate the underlying physics of the planar metamaterial layer without loss of generality, meta-atoms composed of a hexagonal metallic frame or asymmetric double split rings (aDSR) exhibiting Fano-like resonance are periodically arranged (Fig. 1b). Large-area graphene grown by a chemical vapour deposition (CVD) process[26] is transferred onto the meta-atom layer (see Supplementary Information for the characterization of single-layer and multilayer graphene). For gate-controllable doping of the graphene, three electrodes are incorporated into the metamaterial device, two of which are attached on both the top and bottom polyimide spacers surrounding the graphene/meta-atom layer (Fig. 1b), while the other electrode is directly connected to the graphene layer (ground). The top and bottom electrodes are carefully designed to apply a static electric field near the periphery of the graphene layer, while allowing incident terahertz radiation to be transmitted without being perturbed by these electrodes. For this purpose, these electrodes are constructed with a thin metallic wire array, where the gap between adjacent wires is in the deep subwavelength-scale ($\sim\lambda/150$) in order to fully utilize the extraordinary optical transmission (EOT)[27] of terahertz waves (see Supplementary Information for details of the EOT electrodes). By applying the gate voltage to one of the EOT electrodes, the Fermi level of the graphene, and hence the carrier density can be dynamically controlled with a corresponding change of conductivity. This modified optical conductivity is translated into the variation of the complex permittivity of the atomically thin graphene layer, which in turn results in changes to the transmitted terahertz wave through the metamaterial. An optical micrograph of the fabricated graphene metamaterial attached to a printed circuit board (PCB) is shown in Fig. 1c. As shown in Fig. 1d, the fabricated active graphene metamaterial is large-area ($15 \times 15$ mm$^2$), flexible, and free-standing without the thick base substrate that is generally required for semiconductor-based terahertz modulators.

Terahertz time-domain spectroscopy (THz-TDS) was employed to measure amplitude and phase changes in terahertz waves transmitted through the gate-controlled active graphene metamaterial with variations in the applied gate voltage $V_g$. The experimentally measured transmission spectra of the single-layer graphene (SLG) metamaterial (with hexagonal meta-atoms) clearly show the gate voltage dependent resonant features (Fig. 2a): (1) the resonant frequency ($f_0$) is red-shifted with increasing $|\Delta V|$, where $|\Delta V| = |V_{CNP} - V_g|$ and $V_{CNP}$ is the charge neutral gate voltage. This red-shift is mainly due to the increase in the sheet conductivity of graphene; (2) the width of resonance is broadened with increasing $|\Delta V|$, which is caused by additional Joule losses from the *metallic* graphene layer; (3) the on-resonance transmission, $T(f_0)$, increases in proportion to $|\Delta V|^{1/2}$ in the SLG metamaterial as a result of weaker terahertz wave coupling to the resonance of meta-atoms with increasing conductivity in the graphene layer (Fig. 2g); and (4) the off-resonance transmission is supressed with increasing $|\Delta V|$ due to the gate-induced broadband electro-absorption in the graphene layer. The $V_{CNP}$ of the device under test is estimated to be ~350 V from the measured trace of transmission minima (dashed line in Fig. 2a). In order to quantify the active tuning capability more clearly, relative changes in transmission, $-\Delta T/T_{CNP}$ (where $\Delta T = T - T_{CNP}$), are plotted as a function of $V_g$, as shown in Fig. 2b. At the maximum $\Delta V = 850$ V, the measured $-\Delta T/T_{CNP}$ reaches approximately -90 % when the frequency approaches $f_0$ in SLG metamaterial. In the case of a multilayer graphene (MLG) metamaterial, the measured $-\Delta T/T_{CNP}$ exceeds -140 % with a relatively small $\Delta V = 120$ V (see Supplementary Information for the MLG metamaterials). It is noteworthy that this huge modulation is realized solely by the inclusion of a graphene layer of infinitesimal thickness (< $\lambda/1{,}000{,}000$ for SLG) into the metamaterial of deep subwavelength-scale thickness (~$\lambda/100$). The relative change in transmission *per unit thickness* was measured to be 34 %/µm, which exceeds that of all previously reported 2DEG-based terahertz modulators. Moreover, a large degree of *broadband* modulation (more than 10 %/µm) at the off-resonance frequencies (0.1–0.6 and 1.2–2.5 THz) is possible. Figure 2c shows the measured phase change $\Delta\phi$ in SLG metamaterials as a function of $V_g$. The maximum value of $\Delta\phi$ exceeds 40 degrees at 0.65 THz ($\Delta\phi \sim 70$ degrees for the MLG graphene metamaterial, see Supplementary Information). This substantial gate-controlled phase modulation of terahertz waves can be further increased by reducing the gap between the hexagonal meta-atoms and/or by stacking multiple layers of deep subwavelength-scale graphene metamaterial[13].

For a quantitative comparison with experimentally observed phenomena, the transmission spectra of active graphene metamaterials were calculated by a finite element analysis. In order to model the atomically thin graphene layer in electromagnetic wave simulations, the complex conductivity of graphene estimated via Kubo's formula was converted to a dielectric constant for a very thin effective layer[7]. The conductivity of graphene ($\sigma_g = \sigma_{inter} + \sigma_{intra}$) at terahertz frequencies is dominated by the contribution from the intraband transition and can be cast in a Drude-like form. The applied gate voltage leads to a carrier density change and consequently shifts the Fermi level ($E_F$) according to the following formula (for SLG): $E_F = \text{sign}(\Delta V)\hbar v_F(\alpha\pi|\Delta V|)^{1/2}$, where $v_F \approx 1 \times 10^6$ ms$^{-1}$ is the Fermi velocity and $\alpha \approx 5.12 \times 10^9$ cm$^{-2}$V$^{-1}$ is the EOT gate capacitance in the electron charge, respectively. Inserting these values into Kubo's formula, the transmission (Fig. 2d), the relative change in transmission (Fig. 2e), and phase change (Fig. 2f) were calculated as a function of gate voltage changes. All of the gate voltage dependent resonant features were excellently reproduced in the simulations with a single fitting parameter for scattering time $\tau$ = 15 fs, as shown in Fig. 2g. The fitted scattering time is comparable to that previously reported for thermally grown graphene[28]. It is worth noting that the nonlinear relative change in transmission can hardly be fitted with the multilayer approximation[29], which provides additional evidence on the embedded single-layer graphene (single-layer graphene characterization is given in the Supplementary Information). Detailed dynamics of charge carrier transport can be clearly visualized with the numerical simulation. At CNP, the conductivity of the graphene layer is still small enough so as not to perturb the *LC* resonance of hexagonal meta-atoms. With increasing $|\Delta V|$, the graphene becomes more conductive such that fewer charge carriers accumulate at the edge of the meta-atoms as charge carriers can leak into adjacent meta-atoms through conductive graphene channels, as shown in the field simulation depicted in Fig. 2h, i.

To investigate gate-controlled behaviours in an artificial multi-resonance system hybridized with single-layer graphene, asymmetric double split rings (aDSR)[30] are utilized as a unit cell of meta-atoms. Figure 3 shows the gate-dependent transmission for aDSR graphene metamaterial. With broken symmetry in aDSR meta-atoms (asymmetric factors of $\theta_1$ = 15 ° and $\theta_2$ = 10 ° are defined in Fig. 3e), a sharp Fano-like resonance ('trapped mode', marked as I in Fig. 3c) is observed in the low-frequency side of the fundamental electric 'dipole mode' resonance (marked as II in Fig. 3c). The simulated current densities for these two distinct modes in aDSR are depicted by the arrows in the left panel inside Figs. 3e, f. Compared to the electric dipole mode (in-phase current oscillation), the trapped mode resonance (out-of-phase current oscillation) exhibits a higher quality factor due to weak free-space coupling and low radiation losses. At the trapped mode resonance, the absorption of the metamaterial reaches a maximum value of 40 %, the main contribution arising from substrate and metallic losses (the intrinsic terahertz absorption in the graphene layer is substantially lower at CNP[10]). As we increased the gate voltages, absorption increased substantially, except in the vicinity of the Fano-like resonance, by carrier-induced electro-absorption in the single-layer graphene (lower panel of Fig. 3c). Hence, the gate-dependent transmission changes showed different behaviours for these two modes, as exhibited in Fig. 3d. The on-resonance transmission of the trapped mode showed little change ($-\Delta T/T_{CNP} < 5$ %) with gate voltages, whereas for the electric dipole mode, the changes are substantial, amounting to about 40 %. Such broadband tunable absorption means that the graphene metamaterial could be used as a platform to attain a perfect absorber by patterning and/or stacking graphene layers.

In addition to promising amplitude and phase modulation predictable on the basis of the carrier dynamics of *ideal* graphene, active graphene metamaterial exhibits a considerable amount of gate-controlled optical hysteresis. The controllable hysteretic behaviours in the realistic graphene on a substrate[21-24], when combined with the metamaterial, can alternatively be utilized as a route toward the development of an electrically controllable photonic memory. The photonic memory effect in our device can be mainly attributed to charge trapping phenomena at the grain boundaries or defect sites in a large-area CVD-grown graphene layer as well as adsorption of $H_2O$ molecules on the polyimide substrate. Figure 4a shows such hysteretic behaviour in the transmission at a fixed frequency for a cyclic change of the gate voltage (for active graphene metamaterial with hexagonal meta-atoms). For this measurement, the gate voltage was swept at a rate of 50 V/min. Figure 4b shows a flip-flop operation in transmission (top panel) and a time-trace of measured resonant frequency (middle panel). Here, the 'write (set)' and 'erase (reset)' inputs were implemented by applying a short pulse signal of the gate voltage (pulse width = 1 second, lower panel). The peak gate voltage of pulses was set to -300 V (write) and 300 V (erase), respectively. The state of the transmission and/or resonant frequency of the SLG metamaterial was read at zero gate voltage ($V_g = 0$ V, dashed line in Fig. 4a). It is worthwhile to note that the measured retention time of the active graphene metamaterial was proportional to both the magnitude and the pulse width of the applied gate voltage (see Supplementary Information for the results in MLG metamaterial, for which the retention time is larger). With the given gate voltage pulses (±300 V for 1 sec), the retention time is estimated to be around 20 min, which is comparable to the value of CVD-grown SLG ferroelectric memory[24] (Fig. 4b). For longer memory retention times, oxidative graphene would be superior to an intrinsic graphene layer due to oxygen-related, stable memristive phenomena[21,22].

In conclusion, we experimentally demonstrated an electrically controllable light-matter interaction in a gate-controlled active graphene metamaterial. The exotic electrical and optical properties of graphene, when enhanced by the strong resonance of meta-atoms, lead to a very strong light-matter interaction in a manner that allows persistent switching and linear modulation of low energy photons in the extreme subwavelength-scale ($\sim\lambda/1,000,000$). Surprisingly, low energy photons were fully controlled while being transmitted through gate-controlled, one-atom-thick, single-layer graphene integrated with strongly coupled meta-atoms. By stacking the deep subwavelength-scale planar graphene metamaterial, further enhancement of the light-matter interaction is expected. Benefitting from the controllable light-matter interaction, the gate-controlled active graphene metamaterials are expected to provide a myriad of important applications, particularly in the dynamic control of terahertz waves, tunable transformation-optics devices, and photonic memory devices.

**Method Summary**

**Fabrication processes for the gate-controlled active graphene metamaterial** All metallic parts of the graphene metamaterial were made of 100-nm-thick gold with a 10-nm-thick chromium adhesion layer. Graphene was grown by CVD in order to cover the entire area of the metamaterial (15 × 15 mm$^2$). A bare silicon wafer was used as a sacrificial substrate. In order to construct the active graphene metamaterial on the sacrificial substrate, a polyimide solution (PI-2610, HD MicroSystems) was spin-coated and fully cured using a two-step baking process, resulting in a final polyimide thickness of 1 μm. In order to define the bottom EOT electrode on the prepared polyimide layer, UV photolithography and electron-beam evaporation were followed by a metal lift-off technique. Repeating the same polyimide stacking processes as employed for the first layer, the metallic meta-atoms are then insulated from the bottom electrode. The metallic meta-atoms were patterned using the same processes as for the bottom electrode. Then, the graphene (SLG on Cu foil, MLG on Ni/SiO$_2$/Si substrate) was transferred and conformally attached to the array of meta-atoms. After the graphene layer had been transferred onto the meta-atoms, a ground electrode was defined on the graphene layer using a shadow mask. For the symmetry of graphene metamaterials in the direction of the terahertz wave propagation, a polyimide layer and a top EOT electrode (optional) were stacked. After opening the electrical contact via O$_2$ plasma etching, the free-standing and flexible active graphene metamaterials were peeled off of the silicon substrate. Finally, the active graphene metamaterials were soldered to a drilled PCB substrate.

We thank B. H. Hong for the discussion on the application of graphene, Y.-J. Yu for the discussion on carrier transport in graphene, J. H. Han for the characterization of graphene, and H. Choi for proofreading the manuscript. This work was supported by the National Research Foundation of Korea (NRF) grant funded by the Korea government (MEST) (No. 2008-0062235, 2009-0069459, 2010-0012058, 2011-0020186, and 2011-0028151). S. S. L acknowledges the support by the NRF of Korea grant funded by the MEST (No. 2010-0027050). S.-Y. C acknowledges the GFR Program (2011-0031640) sponsored by the MEST. C.-G. C acknowledges the Nano R&D Program (2011-0019169) through the NRF of Korea funded by the MEST and the Creative Research Program of the ETRI (11YF1110). X. Z acknowledges the support from US NSF Nano-scale Science and Engineering Center (NSEC) for Scalable and Integrated Nanomanufacturing (SINAM) at UC Berkeley (Grant No.CMMI-0751621)

Correspondence and requests for materials should be addressed to B. M. (bmin@kaist.ac.kr).

**Figure 1 Schematic view and device images of gate-controlled active graphene metamaterials.** **a**, Schematic rendering of a gate-controlled active graphene metamaterial composed of a single atomic layer of graphene deposited on a layer of hexagonal metallic meta-atoms (a unit cell of $L$ = 60 μm, $g$ = 5 μm, and total device thickness $d$ = 4.2 μm) and top/bottom extraordinary transmission (EOT) electrodes (periodicity = 6 μm, metal width = 4 μm) embedded in a dielectric material of 4.22 μm thickness. The polarization of the incident THz wave is perpendicular to the line electrode, as indicated by the arrows. **b**, Optical micrograph of the fabricated gate-controlled active graphene metamaterial without the top electrode giving a clear view of the hexagonal meta-atoms. **c**, Fully integrated gate-controlled active graphene metamaterial attached to a drilled printed circuit board (PCB) for THz-TDS measurement (B: connected to bottom EOT electrode, G: connected to graphene layer). Inset: Magnified view of the gate-controlled active graphene metamaterial. **d**, Optical image of the fabricated large-area metamaterial wound around a glass rod, showing its high degree of flexibility.

**Figure 2. Gate-controlled amplitude and phase changes of terahertz waves transmitted through the hexagonal graphene metamaterials** Measured spectra of **a**, transmission ($T$), **b**, relative change in transmission (-$\Delta T/T_{CNP}$), **c**, phase change ($\Delta\phi$) plotted as a function of gate voltages. **d**, **e**, **f**, Simulated results with a single layer graphene (SLG) approximation are plotted, corresponding to a, b, and c, respectively. A fitting parameter for scattering time $\tau$ is set to 15 fs. **g**, The relative change in transmission at a resonance frequency of 0.86 THz is plotted. Scatters and lines are for the experimental and simulation data, respectively. The measured nonlinear trend in transmission change is

excellently fitted by the SLG approximation.  **h**, Saturated electric field (at 0.86 THz) within a unit cell for the graphene metamaterial with (lower panel) and without applying gate voltage (upper panel).  **i**, Current density (at 0.86 THz) in the meta-atom (left panel) and the graphene layer (right panel) with increasing gate voltages.

**Figure 3. Gate-controlled amplitude change of terahertz waves transmitted through the asymmetric double split ring (aDSR) graphene metamaterials a**, Measured relative change in transmission of the aDSR graphene metamaterials is plotted as a function of gate voltage.  **b**, Simulated results obtained with the SLG approximation show good agreement with experimental observation.  **c**, Two distinct resonances are marked I (trapped mode) and II (dipole mode) with different tuning behaviours.  As the gate voltage increased, transmission at the dipole mode resonance is significantly increased; however, the trapped mode transmission showed little change, where the origin is from a small change in the absorption for the trapped mode (lower panel).  **d**, Relative changes in transmission for the two modes are plotted along with the simulations.  **e**, Gate-dependent current density for the trapped mode is shown with the arrows in a unit cell of aDSR meta-atom (left panel) and graphene layer (right panel).  **f**, Corresponding current density for the dipole mode.

**Figure 4. Electrically-controlled photonic memory operation with the gate-controlled active graphene metamaterials**   **a**, Hysteresis behaviour of on-resonance transmission for a cyclic change of the gate voltage (with a sweeping rate of 50 V/min).  **b**, Binary memory operation in the transmission (top panel), a measured time-trace of resonant frequency (middle panel), and gating pulse

signal (pulse width = 1 sec, lower panel). The observed memory retention time is estimated to be around 20 min.

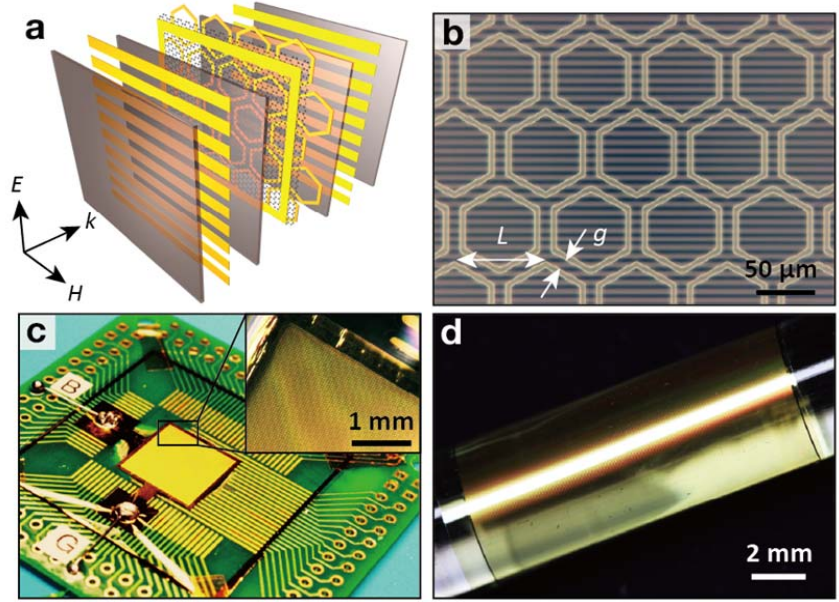

**Lee, Figure 1**

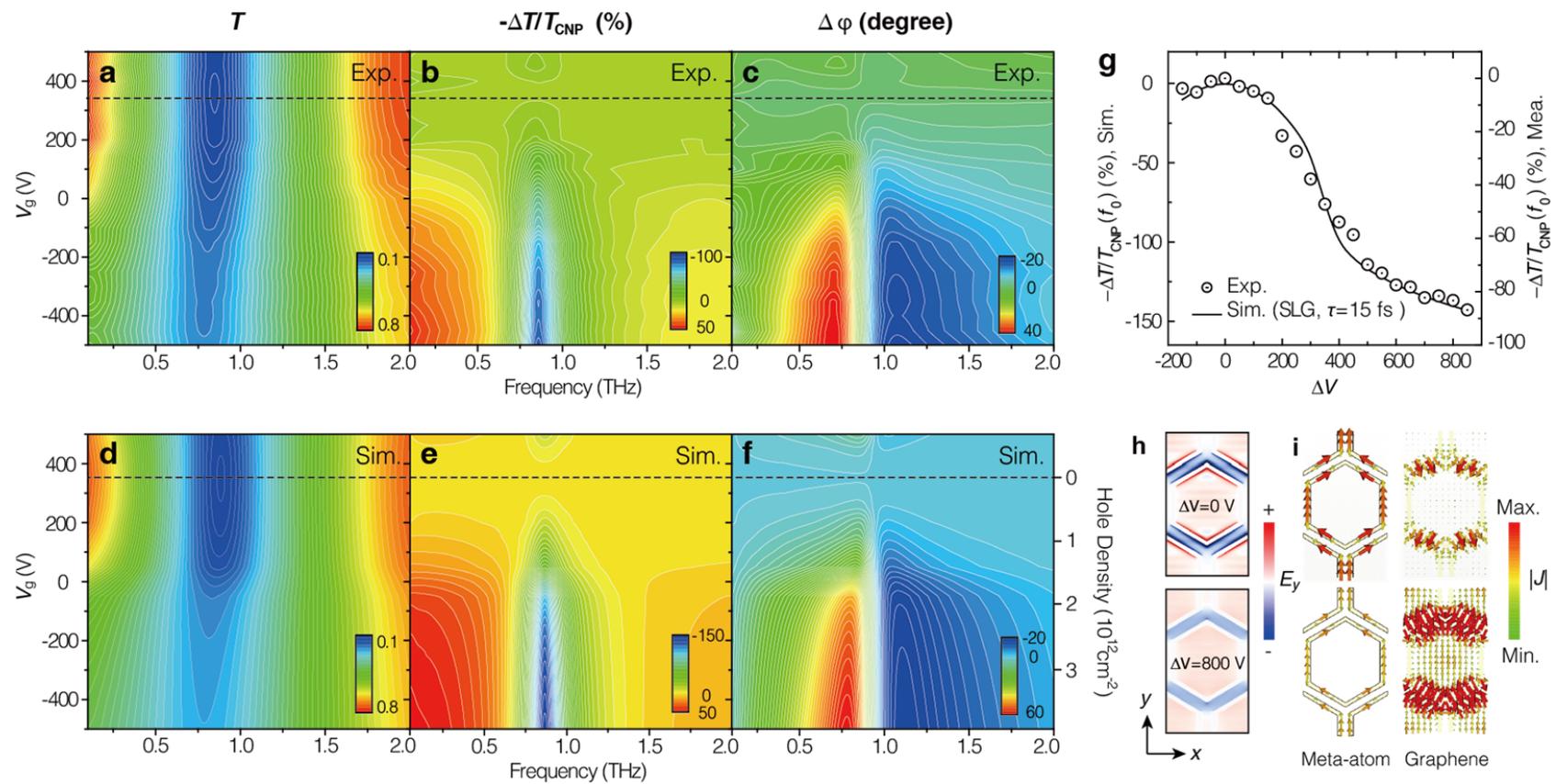

Lee, Figure 2

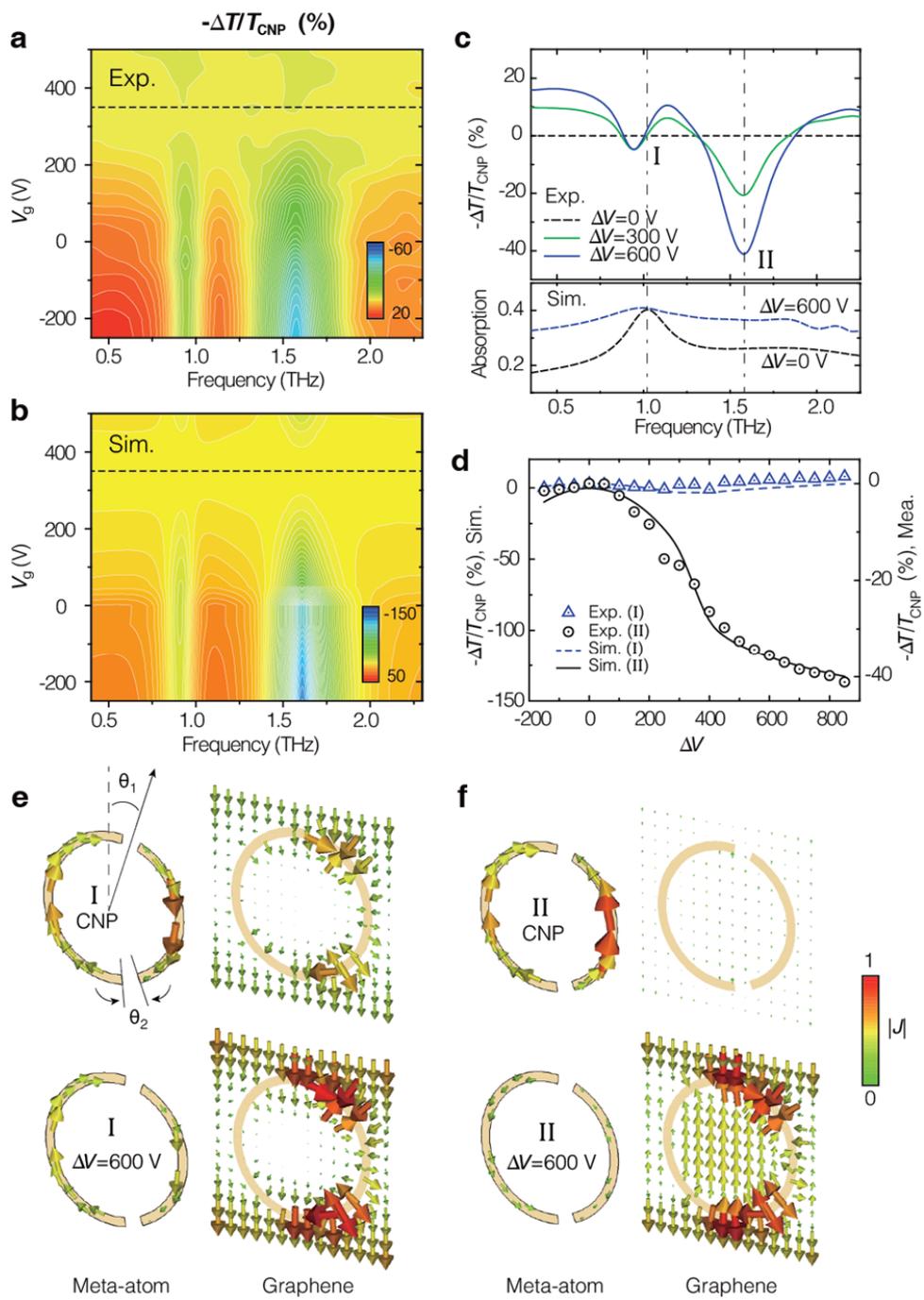

**Lee, Figure 3**

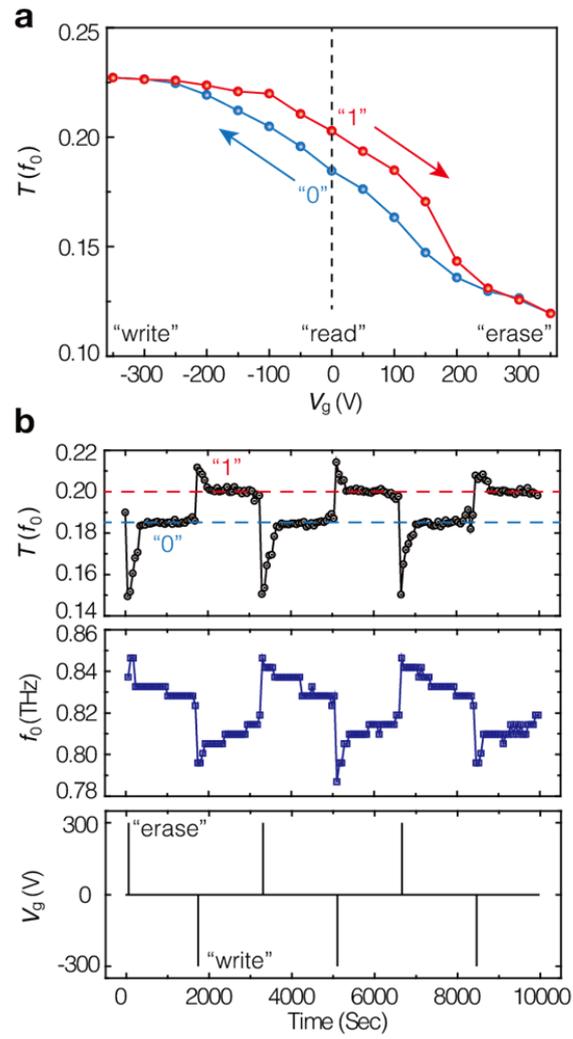

**Lee, Figure 4**

# Supplementary Information for "Switching terahertz waves with gate-controlled active graphene metamaterials"

Seung Hoon Lee, Muhan Choi, Teun-Teun Kim, Seungwoo Lee, Ming Liu, Xiaobo Yin, Hong Kyw Choi, Seung S. Lee, Choon-Gi Choi, Sung-Yool Choi, Xiang Zhang & Bumki Min

Here, we present the preparation and characterization of graphene, the design of extraordinary optical transmission (EOT) gate electrodes, the experimental/numerical data on active multilayer graphene (MLG) metamaterials, and the results of control experiments on samples without graphene.

I.     **Preparation and characterization of graphene**

    A.     **CVD growth of SLG and MLG and their transfer method**

    B.     **Characterization of CVD-grown graphene**

II.    **Design of EOT gate electrodes**

III.   **Terahertz wave transmission properties of active MLG metamaterials**

    A.     **Amplitude modulation**

    B.     **Phase modulation**

    C.     **Hysteretic behaviours**

IV.   **Measurement of control samples without graphene layer**

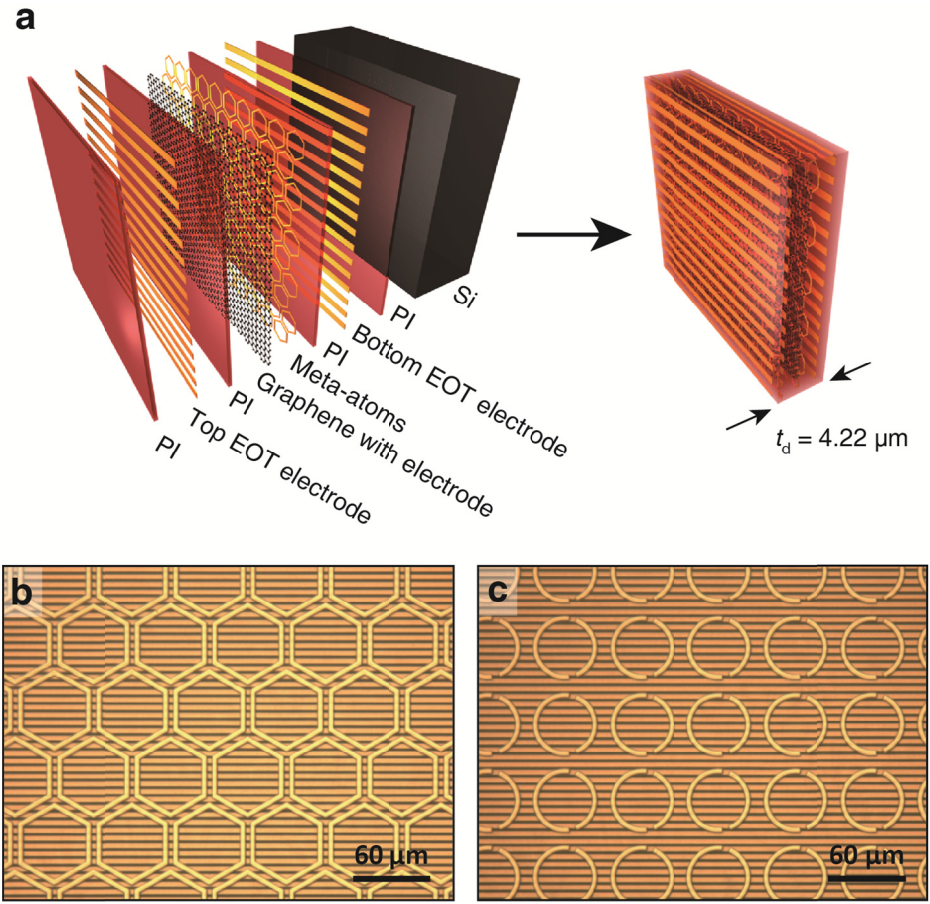

**Figure S1 | Schematics of the gate-controlled active graphene metamaterials and their optical images a**, Schematic view on fabrication of the active graphene metamaterial (PI: polyimide, $t_d$: total device thickness). **b**, Optical micrographs of the fabricated hexagonal meta-atoms **c**, Asymmetric double split ring meta-atoms.

## I. Preparation and characterization of graphene

### A. CVD growth of SLG and MLG and their transfer method

For the fabrication of samples, we used commercially available single-layer graphene (SLG) from Graphene Square Inc., and multilayer graphene (MLG) grown by in-house thermal chemical vapour deposition (CVD). For MLG growth, 300-nm-thick nickel was deposited as a catalyst layer on a $SiO_2$/Si substrate by an electron-beam evaporator. Prepared Ni/$SiO_2$/Si substrate was loaded in a quartz tube furnace, then heated up to 1,000 °C under vacuum condition (30 mTorr) with a constant $H_2$ flow (10 sccm). MLG was grown by flowing gas mixtures ($CH_4$:$H_2$ / 30:10 sccm) for 3 min and MLG on the substrate was dismounted from the furnace and cooled down to room temperature (25 °C) at a cooling rate of 5 °C/min under Ar gas flow of 1,000 sccm.

In order to transfer the CVD-grown SLG and MLG to a polyimide substrate, a polydimethylsiloxane (PDMS) film was used to minimize undesired cracks and solvent-induced effects. The thermally grown SLG (MLG) on the Cu foil (Ni/$SiO_2$/Si substrate) was peeled off from the substrate after 8-hour etching in a 0.1 mol% ammonium persulfate (3-hour etching in a 12.5 mol% iron chloride) solution. After a two-step rinsing process with DI water, the graphene layer was successfully transferred and conformally attached to the prepared polyimide layer, on which the meta-atoms were patterned.

### B. Characterization of CVD-grown graphene

Raman spectroscopy was used with a 532 nm excitation laser for the characterization of graphene layers. The SLG and MLG, grown in the same batch for samples used in the main manuscript, were transferred onto a 300-nm-thick $SiO_2$ substrate. Based on 2D/G intensity ratios and the full width at half maximum (FWHM), the number of graphene layers was estimated. As shown in Fig. S2a, the location and FWHM of 2D peak is 2,686 $cm^{-1}$ and 30 $cm^{-1}$, respectively. For MLG (Figs. S2b-d), the number of layer varies spatially and is estimated to be approximately 1 to 10, as confirmed by transmission electron micrographs (TEM). A cross-sectional TEM image of the fabricated MLG metamaterial shows that there are several discontinuities with defects sites, such as gaps and wrinkles, which can be attributed to the one of the dominant causes of electrically controlled photonic memory effects.

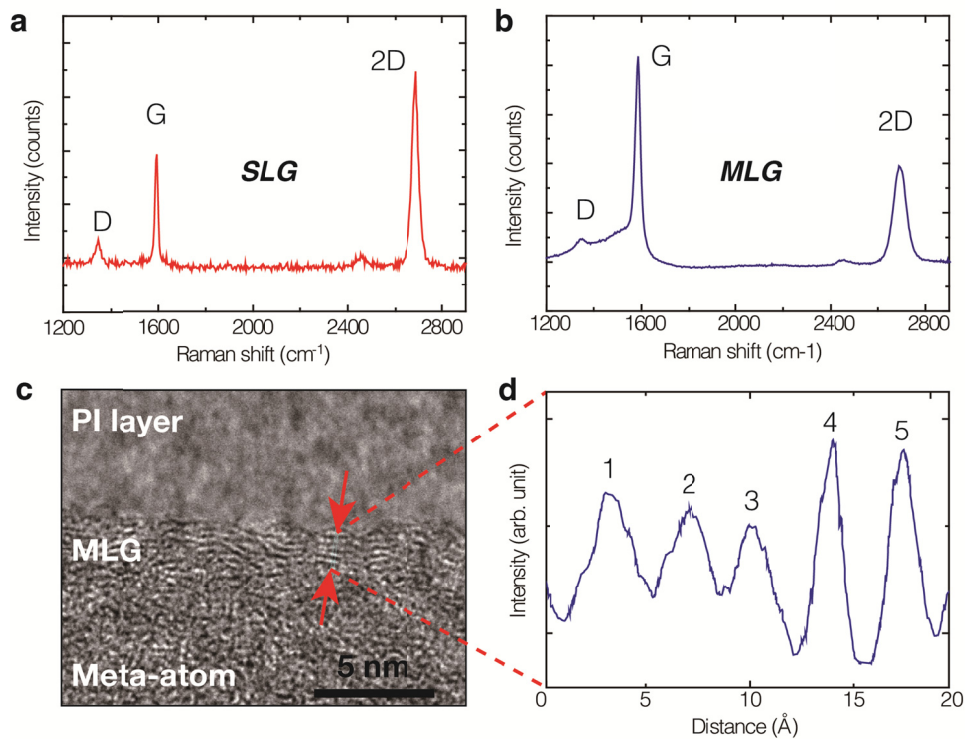

**Figure S2 | Characterization of CVD-grown graphene by Raman spectroscopy and transmission electron microscopy (TEM). a**, Raman spectrum of SLG shows that the 2D peak is located at 2,686 cm$^{-1}$ and the full width at half maximum (FWHM) is about 30 cm$^{-1}$. **b**, Raman spectrum of MLG shows that the FWMH of 2D peak is larger than 45 cm$^{-1}$. **c**, A cross-sectional TEM image of the active MLG metamaterial. Between the polyimide layer and metallic meta-atoms, layered graphene is clearly identified. **d**, Average intensity profile across the graphene layer marked by red arrows.

## II. Design of EOT gate electrodes

The gate electrode for the active graphene metamaterial was designed to apply a uniform electric field in the graphene layer while maintaining a large transmission of terahertz waves. In order to satisfy this requirement, an array of deep subwavelength-scale metallic wires, referred to as an EOT electrode, was optimized. Figure S3a shows the schematics (inset) of the EOT electrode dimension with numerically calculated transmission at terahertz frequencies (0 - 2 THz). The EOT electrodes for the sample were fabricated with the periodicity ($a$) of 6 μm and the gap width ($a - w$) of 2 μm. Although the gap width of the wire array is in the deep subwavelength-scale (~$\lambda/150$ at 1 THz), the simulated transmission is found to be over 89 per cent in the frequency band of interest. This large transmission can be explained on the basis of well-known phenomena of extraordinary optical transmission[1]. Figure S3b shows the calculated and measured transmission through the active graphene metamaterial with EOT electrodes.

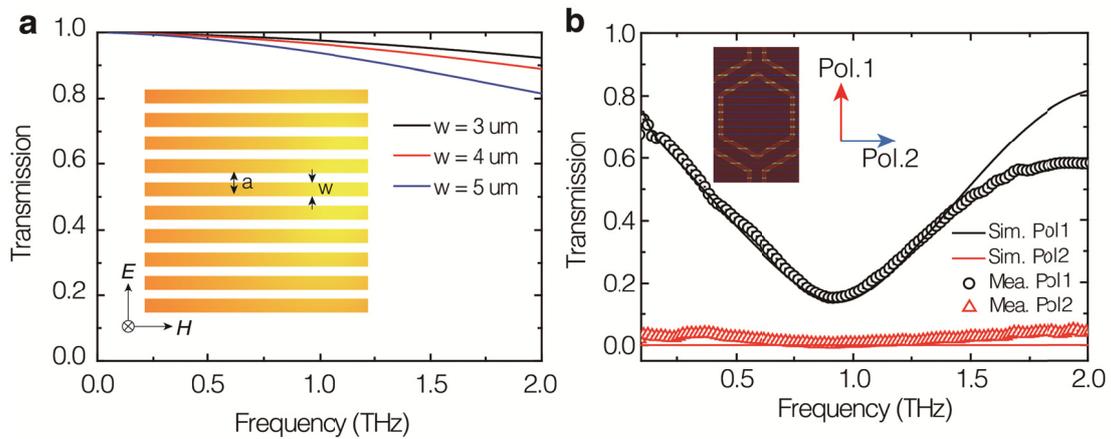

**Figure S3 | EOT electrodes for the gate-controlled active graphene metamaterials. a**, Schematic rendering of EOT electrodes along with simulated transmission spectra for wire widths of 3, 4, and 5 μm. **b**, Polarization dependent transmission through the active graphene metamaterials integrated with EOT electrodes.

## III. Terahertz wave transmission properties of active MLG metamaterials

### A. Amplitude modulation

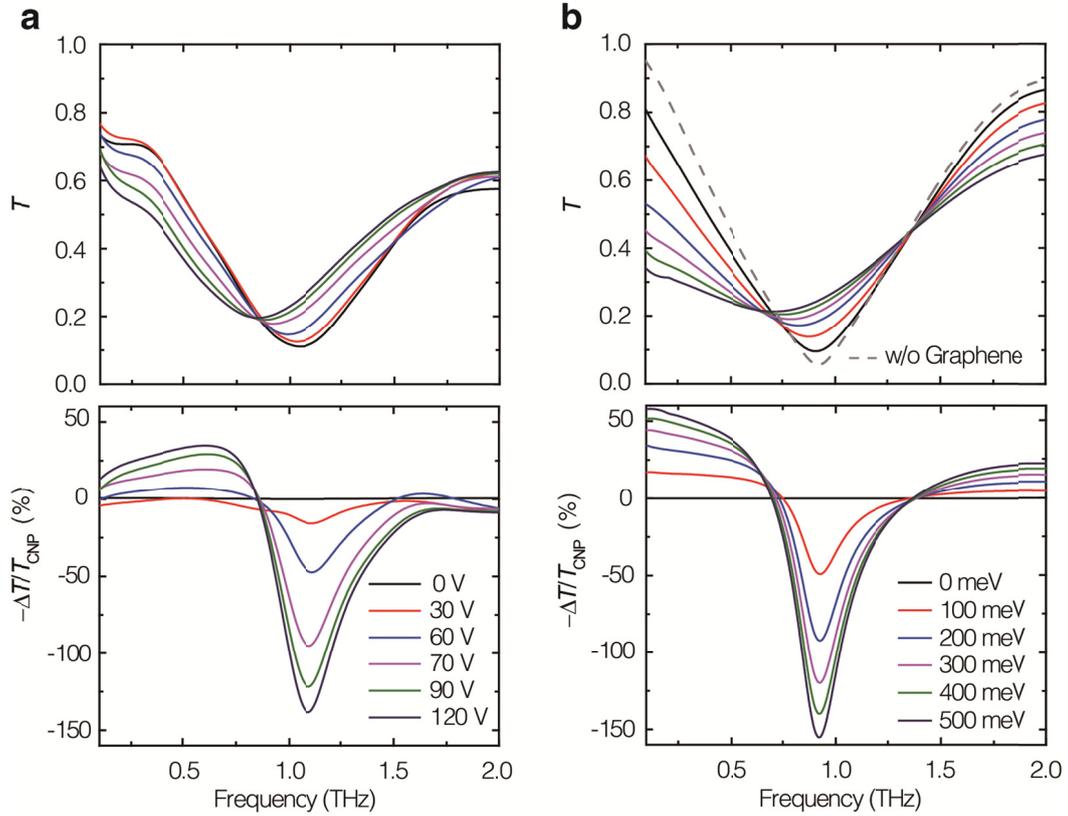

**Figure S4 | Amplitude modulation of terahertz waves with the gate-controlled active MLG metamaterial.** **a**, Measured transmission spectra as a function of $\Delta V$ (= $V_{CNP} - V_g$) from 0 to 120 V. With increasing gate voltage, the resonant frequency was red-shifted and broadened by Joule losses in the graphene layer. Relative change in transmisstion, $-\Delta T/T_{CNP}$ exceeded -140 % at 120 V. **b**, For comparison, the simulated transmission of the active graphene metamaterials are plotted as a function of the Fermi level in a graphene layer (Here, we assumed the intraband scattering time of 25 fs).

## B. Phase modulation

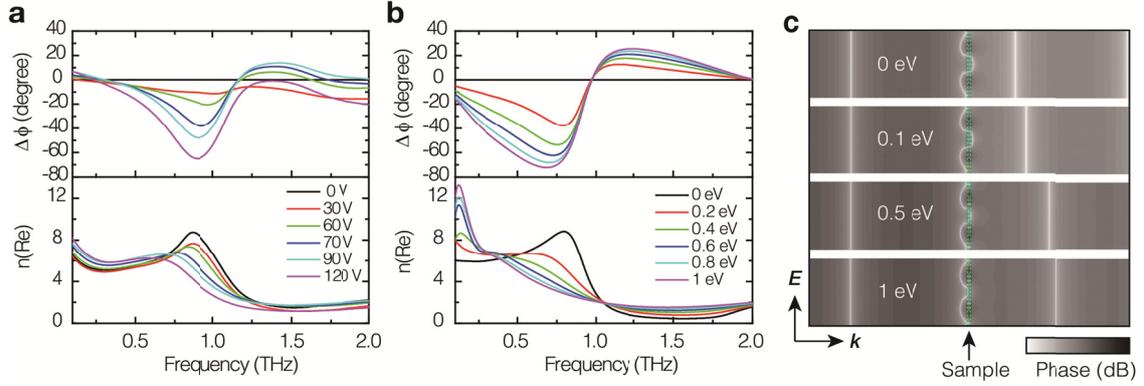

**Figure S5 | Phase modulation of terahertz waves with the gate-controlled MLG metamaterials.** **a**, Measured phase change of terahertz waves through the MLG metamaterials as a function of $\Delta V$. Maximum phase change exceeded $\pi/3$ radians at $\Delta V$ = 120 V. The effective refractive indices were estimated from the S-parameter extraction method and were shown in the lower panels. The gate-controlled refractive index was varied from 4 to 9 with increasing gate voltage at resonant frequency. **b**, Simulated phase change of terahertz waves through the active graphene metamaterial as a function of Fermi energy level. **c**, Simulated wavefronts of terahertz wave are shown as a function of Fermi level at a frequency of 0.64 THz. Phase delays are clearly observed with increases in the Fermi level.

## C. Hysteretic behaviours

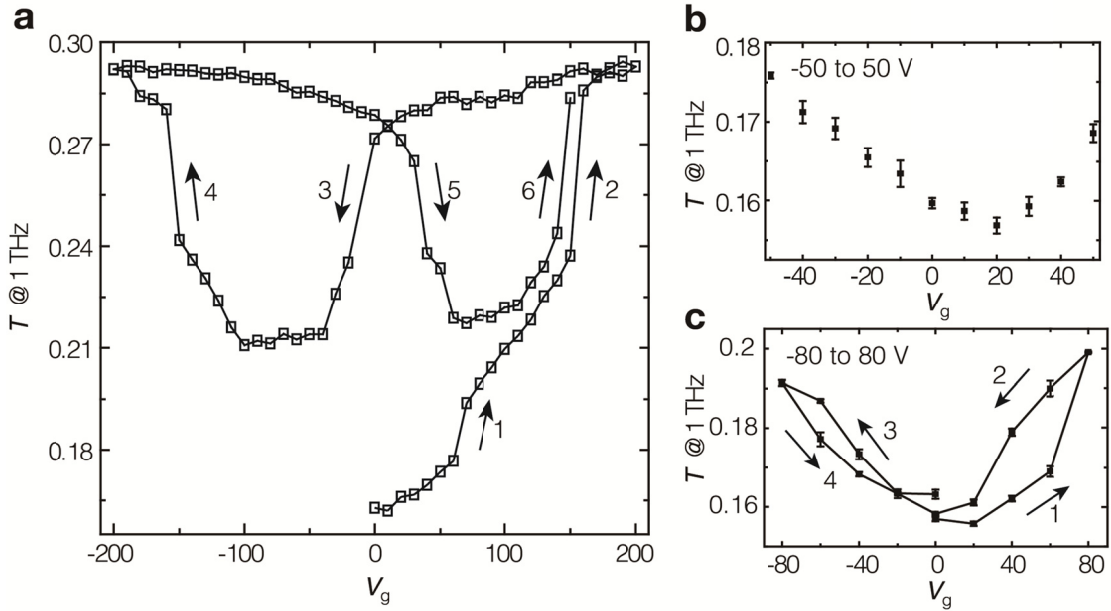

**Figure S6 | Gate-dependent optical hysteresis of MLG metamaterials. a**, Hysteresis curve of the gate-controlled active graphene metamaterial ($V_{CNP} \approx 20$ V, gate voltage sweeping rate of ~3 V/min, transmission changes following the path indicated by arrows). The hysteresis curve was dependent upon the gate voltage sweeping rate and memory retention time. **b**, For small gate voltage change near CNP (from -50 to 50 V), transmission measured at 1 THz was varied linearly with little hysteresis. **c**, For large gate voltage change between -80 to 80 V, a hysteretic behaviour was clearly observed.

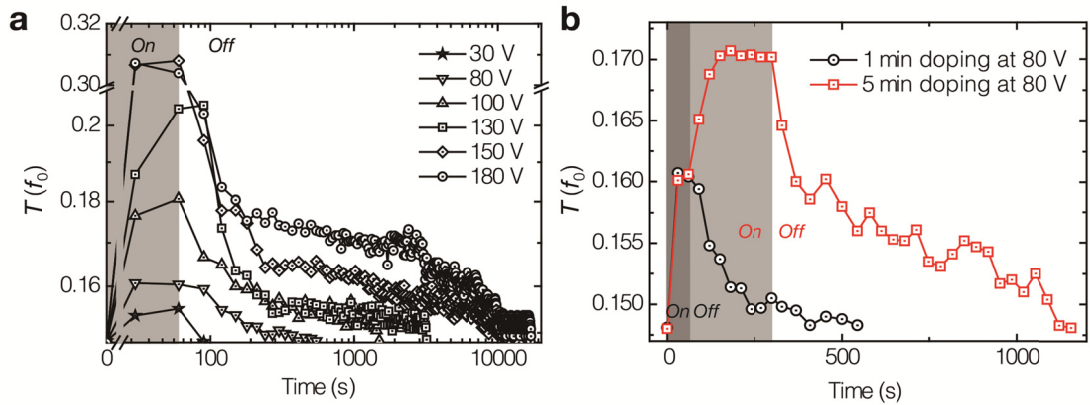

**Figure S7 | Measured retention time of the MLG metamaterials.** **a**, Measured retention time as a function of gate voltage. An initial on-resonance transmission (0.148) at 0 V is shifted when applying gate voltage to 0.307 at 150 V. After 1 min doping at various gate voltages, the transmission was recovered to the initial values. Within small gate voltage at 30 V, the resonance recovery time was fast enough as not to be measured using our THz-TDS setup. **b**, Doping time dependent recovery time at a gate voltage of 80 V. As the electro-static doping time (gating time) is increased, the retention time also increased.

## IV. Measurement of control samples without graphene layer

In order to confirm the effect of gate-controlled graphene in active graphene metamaterials, two graphene-free metamaterials with different gaps (5 um and 8 um) were fabricated and their transmission spectra were measured with a variation in gate voltages. As can be seen from Fig. S8, the measured transmission spectra were almost invariant to the change in gate voltage applied to the top and bottom EOT electrodes.

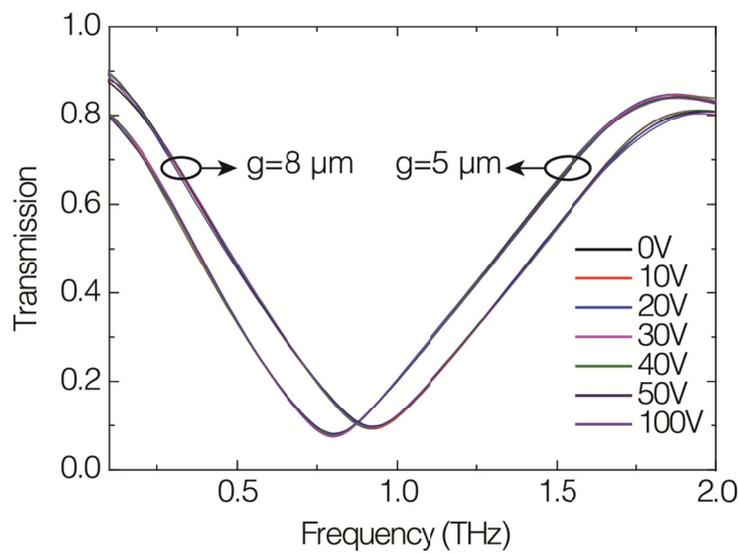

**Figure S8 | Measured transmission spectra of the metamaterials without graphene layer.** The gate voltage was varied from 0 to 100V.

**References**

1  Ebbesen, T. W., Lezec, H. J., Ghaemi, H. F., Thio, T. & Wolff, P. A. Extraordinary optical transmission through sub-wavelength hole arrays. *Nature* **391**, 667-669 (1998).